\documentclass[twocolumn,showpacs,preprintnumbers,amsmath,amssymb]{revtex4}
\usepackage{graphicx}
\usepackage{dcolumn}
\usepackage{bm}
\usepackage{epsf}
\newcommand{\p}{\partial}
\newcommand{\omu}{\overline{\mu}}
\newcommand{\om}{\overline{m}}
\newcommand{\og}{\overline{g}}
\newcommand{\oD}{\overline{\Delta}}
\newcommand{\de}{\delta}
\newcommand{\De}{\Delta}
\newcommand{\f}{\frac}
\newcommand{\lms}{\Lambda_{\overline{MS}}}
\newcommand{\obeta}{\overline{\beta}}
\newcommand{\ogamma}{\overline{\gamma}}
\newcommand{\okappa}{\overline{\kappa}}
\newcommand{\ms}{\overline{MS}}
\newcommand{\B}{\beta}
\newcommand{\C}{\gamma}
\begin{document}
\preprint{RUG preprint}
\title{The mass gap and vacuum energy of \\the Gross-Neveu model via the $2PPI$ expansion}
\author{David Dudal}
    \altaffiliation{Research Assistant of the Fund For Scientific
Research-Flanders (Belgium)}
    \email{david.dudal@rug.ac.be}
\author{Henri Verschelde}
 \email{henri.verschelde@rug.ac.be}
\affiliation{Ghent University
\\ Department of Mathematical
Physics and Astronomy \\ Krijgslaan 281-S9 \\ B-9000 GENT,
BELGIUM}
\begin{abstract}
We introduce the $2PPI$ (2-point-particle-irreducible) expansion,
which sums bubble graphs to all orders. We prove the
renormalizibility of this summation. We use it on the Gross-Neveu
model to calculate the mass gap and vacuum energy. After an
optimization of the expansion, the final results are qualitatively
good.
\end{abstract}
 \maketitle
\section{\label{sec1}Introduction}
The Gross-Neveu (GN) model \cite{Gross} is plagued by infrared
renormalons. The origin of this problem lies in the fact that we
perturb around an instable (zero) vacuum. A remedy would be the
mass generation of the particles, connected to a non-perturbative,
lower value of the vacuum energy. Such a dynamical mass must be of
a non-perturbative nature, since the GN Lagrangian possesses a
discrete chiral symmetry. A dynamical mass is closely related to a
nonzero vacuum expectation value (VEV) for a local composite
operator $\left(\textrm{i.e. }\overline{\psi}\psi\right)$. This
condensate introduces a mass scale into the model. We consider GN
because the exact mass gap \cite{Forgacs} and vacuum energy
\cite{Zamo} are known. This allows a test for the reliability of
approximative frameworks before attention is paid to dynamical
mass generation in more complex theories like SU($N$) Yang-Mills
\cite{Verschelde1}. The last few years, several methods have been
proposed to solve this problem and get non-perturbative
information out of the model
\cite{Arvanitis,Verschelde3,Karel}.\\In this paper, we address
another approach, the so-called $2PPI$ expansion. Its first
appearance and use for analytical finite temperature research can
be found in
\cite{Verschelde4,Verschelde5,Verschelde6,Verschelde7,Baacke}. In
Sec.\ref{sec2}, we give a new derivation of the expansion.
Sec.\ref{sec3} is devoted to the renormalization of the $2PPI$
technique. Preliminary numerical results, using the $\ms$ scheme,
are presented in Sec.\ref{sec4}. We recover the
$N\rightarrow\infty$ approximation, but we encounter the problem
that the coupling is infinite. In Sec.\ref{sec5} we optimize the
$2PPI$ technique. We rewrite the expansion in terms of a scheme
and scale independent mass parameter $M$. The freedom in coupling
constant renormalization is reduced to a single parameter $b_{0}$
by a reorganization of the series. We discuss how to fix $b_{0}$.
Numerical results can be found in Sec.\ref{sec6}. We also give
some evidence to motivate why results are acceptable. We end with
conclusions in Sec.\ref{sec7}.
\section{\label{sec2}The $2PPI$ expansion}
We start from the (unrenormalized) GN Lagrangian in
two-dimensional Euclidean space time.
\begin{equation}\label{1}
    \mathcal{L}=\overline{\psi}\partial\hspace{-0.2cm}/\psi-\frac{1}{2}g^{2}\left(\overline{\psi}\psi\right)^{2}
\end{equation}
This Lagrangian has a global $U(N)$ invariance and a discrete
chiral symmetry $\psi\rightarrow\C_{5}\psi$ which imposes
$\left\langle\overline{\psi}\psi\right\rangle=0$ perturbatively.
This model is asymptotically free and has spontaneous chiral
symmetry breaking. As such, it is a toy model which mimics QCD in
some ways.\\First of all, we focus on the topology of vacuum
diagrams. We can divide them in 2 disjoint classes:
\begin{itemize}
    \item Those diagrams falling apart in 2 separate
    pieces when 2 lines meeting at the same point $x$ are cut. We call those \textit{2-point-particle-reducible} or $2PPR$. $x$ is named the
    $2PPR$ insertion point. FIG.1 depicts the most simple $2PPR$ vacuum
    bubble.
\begin{figure}[htb]\label{fig1}
\begin{center}
    \scalebox{0.6}{\includegraphics{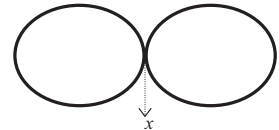}}
    \caption{A $2PPR$ vacuum bubble. $x$ is the $2PPR$ insertion point.}
\end{center}
\end{figure}
    \item The other type is the complement of the $2PPR$ class, we
    baptize such diagrams \textit{2-point-particle-irreducible} ($2PPI$) diagrams.
    FIG.2 shows a $2PPI$ bubble.
\begin{figure}[htb]\label{fig2}
\begin{center}
    \scalebox{0.6}{\includegraphics{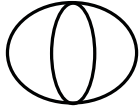}}
    \caption{A $2PPI$ vacuum bubble.}
\end{center}
\end{figure}
\end{itemize}
We could now remove all $2PPR$ bubbles from the diagrammatic sum
building up the vacuum energy by summing them in an effective
mass. To proceed, we must use a little trick. Let's define
\begin{equation}\label{2}
    \Delta = \langle\overline{\psi}\psi\rangle = \left\langle\overline{\psi}_{i}\psi_{i}\right\rangle
\end{equation}
where the index $i=1\ldots 2N$ goes over space as well as internal
values. Obviously, we have
\begin{equation}\label{2bis}
    \Delta_{ij}\equiv\left\langle\overline{\psi}_{i}\psi_{j}\right\rangle=\delta_{ij}\frac{\Delta}{2N}
\end{equation}
We now calculate $\f{dE}{d g^{2}}$ where $E$ is the vacuum energy.
The $g^{2}$ derivative can hit a $2PPR$ vertex or a $2PPI$ vertex.
(see FIG.3 and FIG.4)
\begin{figure}[t]\label{fig3}
\begin{center}
    \scalebox{0.6}{\includegraphics{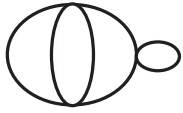}}
    \caption{Generic vacuum bubble.}
\end{center}
\end{figure}
\begin{figure}[t]\label{fig4}
\begin{center}
    \scalebox{0.6}{\includegraphics{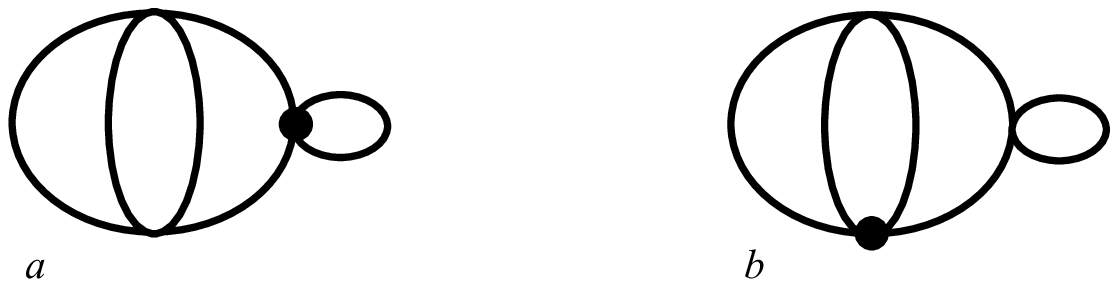}}
    \caption{Diagrammatic depiction of $\frac{d}{dg^{2}}$ (fat dot) applied on the bubble of FIG.4.}
\end{center}
\end{figure}
In the first case, we have diagrammatically the contribution
\begin{equation}\label{eq1}
    -\f{1}{2}\De_{ij}\left(\de_{ij}\de_{kl}-\de_{il}\de_{jk}\right)\De_{kl}=-\f{1}{2}\left(1-\f{1}{2}\right)\De^{2}
\end{equation}
In the second case, we can unambiguously subdivide the vacuum
diagram in one maximal $2PPI$ part, which contains the vertex hit
by $\f{d}{d g^{2}}$, and one or several $2PPR$ parts which can be
deleted and replaced by an effective mass $\om$. A simple
diagrammatical argument gives
\begin{equation}\label{eq2}
    \om\de_{ij}=-g^{2}\left(\de_{ij}\de_{kl}-\de_{il}\de_{jk}\right)\De_{kl}
\end{equation}
or
\begin{equation}\label{eq3}
    \om=-g^{2}\Delta\left(1-\frac{1}{2N}\right)
\end{equation}
Summarizing, we have
\begin{eqnarray}\label{eq4}
    \frac{dE}{d g^{2}}
    =-\frac{1}{2}\Delta^{2}\left(1-\frac{1}{2N}\right)+\frac{\partial E_{2PPI}}{\partial g^{2}}\left(\om,g^{2}\right)
\end{eqnarray}
The $g^{2}$ dependence in $E_{2PPI}$ comes from the $2PPI$
vertices. To integrate (\ref{eq4}), we use the Anzatz
\begin{equation}\label{eq5}
    E\left(g^{2}\right) = E_{2PPI}\left(\om,g^{2}\right) + cg^{2}\Delta^{2}
\end{equation}
with $c$ a constant to be determined. Using (\ref{eq3}), we find
from (\ref{eq5})
\begin{eqnarray}\label{eq6}
    \frac{dE}{d g^{2}} &=& \frac{\partial E_{2PPI}}{\partial g^{2}}\left(\om,g^{2}\right) +
    \frac{\partial E_{2PPI}}{\partial \om}\left(-\Delta\left(1-\frac{1}{2N}\right)\right.\nonumber\\&-&\left.
    g^{2}\frac{d
    \Delta}{dg^{2}}\left(1-\frac{1}{2N}\right)\right)+c\Delta^{2}+2cg^{2}\Delta\frac{d\Delta}{dg^{2}}
\end{eqnarray}
A simple diagrammatical argument gives
\begin{equation}\label{eq7}
    \frac{\partial E_{2PPI}}{\partial \om}\left(\om,g^{2}\right)=\Delta
\end{equation}
This is a (local) gap equation, summing the bubble graphs into
$\om$. Using (\ref{eq7}) and comparing (\ref{eq6}) with
(\ref{eq4}), we find $c=\frac{1}{2}\left(1-\frac{1}{2N}\right)$,
so that we finally have that
\begin{equation}\label{eq8}
    E\left(g^{2}\right)=\f{1}{2}g^{2}\left(1-\f{1}{2N}\right)\De^{2}+E_{2PPI}\left(\om,g^{2}\right)
\end{equation}
It is easy to show that the following equivalence hold.
\begin{equation}\label{tra}
    \f{\p E_{2PPI}}{\p \om}=\De \Leftrightarrow \f{\p E}{\p \om}=0
\end{equation}
One shouldn't confuse (\ref{tra}) with the usual procedure of
minimizing an effective potential $V(\varphi)$ with respect to the
field variable $\varphi$. First of all, $\om$ is not a field
variable. Secondly, the expression for $E$ in terms of the $2PPI$
expansion is only correct if the gap equation is fulfilled.
\section{\label{sec3}Renormalization of the $2PPI$ expansion}
Up to now, we haven't paid any attention to divergences. We will
now show that an equation such as (\ref{eq8}) is valid for the
vacuum energy $E$ with fully renormalized and finite quantities.
Since in the original Lagrangian there is no mass counterterm, one
could naively expect problems with the non-perturbative mass
$\om$, which generates mass renormalization in $E_{2PPI}$. Another
possible problem is vacuum energy renormalization. Perturbatively,
the vacuum energy is zero and hence no vacuum energy
renormalization is needed. Non-perturbatively, we expect
logarithmic divergences proportional to $\om^{2}$ for $E_{2PPI}$.
As we will show, both these problems are solved with coupling
constant renormalization.\\\\The trick is to separate the
contribution of the coupling constant renormalization counterterm
$-\f{1}{2}\de Z_{4}\left(\overline{\psi}\psi\right)^2$ into $2PPR$
and $2PPI$ parts, corresponding with the topology of the original
divergent subgraphs. Let $i$ and $j$ be the indices carried by the
lines meeting at the $2PPR$ vertex, then we have
\begin{equation}\label{R11}
    \delta
    Z_{4}\left(\de_{ij}\de_{kl}-\de_{il}\de_{kj}\right)=\de
    Z_{4;ij,kl}^{2PPI}+\de
    Z_{4;ij,kl}^{2PPR}
\end{equation}
Note that crossing will change a $2PPR$ part into a $2PPI$ part.
\begin{figure}[t]\label{fig5}
\begin{center}
    \scalebox{0.5}{\includegraphics{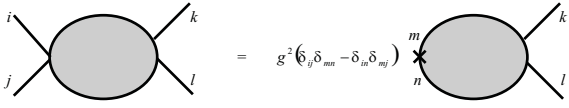}}
    \caption{A diagrammatical identity.}
\end{center}
\end{figure}
\\\\
Because of the diagrammatical identity shown in FIG.5, with
$^{m}\textrm{---x---}^{n}$ a $\overline{\psi}_{m}\psi_{n}$
insertion, we have a relation between the $2PPR$ part of coupling
constant and mass renormalization.
\begin{equation}\label{R12}
    \de
    Z_{4;ij,kl}^{2PPR}=\left(\de_{ij}\de_{mn}-\de_{in}\de_{mj}\right)\de
    Z_{2;mn,kl}
\end{equation}
This identity can be used to show that the divergent effective
mass $\om$, given by (\ref{eq3}), gets replaced by a finite
renormalized mass
$\om_{R}=Z_{2}\om=-g^{2}\left(1-\f{1}{2N}\right)\De_{R}$, where
$\De_{R}=Z_{2}\De$ is the finite, renormalized expectation value
of the composite operator $\overline{\psi}\psi$. Indeed, let us
consider a generic $2PPR$ subgraph or bubble graph with a $2PPR$
vertex $x$. The divergent subgraphs of this bubble graph, which do
not contain $x$, can be made finite by the usual counterterms for
wavefunction and coupling constant renormalization. The resulting
effective mass will be given by (\ref{eq3}), but now with
$\De=\left\langle\overline{\psi}{\psi}\right\rangle$ evaluated
with the full Lagrangian, i.e. including counterterms. We still
have to consider the subgraphs of the bubble graph which do
contain the $2PPR$ vertex $x$. They can be made finite by coupling
constant renormalization, but because the subgraph is $2PPR$ at
$x$, only the $2PPR$ part of the counterterm has to be inserted
and we get the contribution (see FIG.6)
\begin{equation}\label{R13}
    -g^{2}\de Z_{4;ij,kl}^{2PPR}\De_{kl}=-g^{2}\left(\de_{ij}\de Z_{2;mm,kk}-\de
    Z_{2;ij,kk}\right)\f{\De}{2N}
\end{equation}
where use was made of (\ref{2bis}) and (\ref{R12}).
\begin{figure}[t]\label{fig6}
\begin{center}
    \scalebox{0.5}{\includegraphics{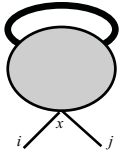}}
    \caption{Divergent subgraph containing the $2PPR$ vertex $x$. Fat lines denote full propagators. }
\end{center}
\end{figure}
Since for a diagonal mass matrix, we can define $\de Z_{2}$ by
\begin{equation}\label{R14}
    \de Z_{2}\de_{kl}=\de Z_{2;mm,kl}
\end{equation}
we have
\begin{equation}\label{R15}
    2N\de Z_{2}=\de Z_{2;mm,kk}
\end{equation}
and
\begin{equation}\label{R16}
    \de Z_{2;ij,kk}=\de Z_{2}\de_{ij}
\end{equation}
After substition of (\ref{R15}) and (\ref{R16}) into (\ref{R13}),
we find that the contribution of the $2PPR$ counterterm insertion
gives
\begin{equation}\label{R17}
    -\de_{ij}g^{2}\left(1-\f{1}{2N}\right)\de Z_{2}\Delta
\end{equation}
and hence a mass renormalization
\begin{equation}\label{R18}
    \de \om=-g^{2}\left(1-\f{1}{2N}\right)\de Z_{2}\Delta
\end{equation}
so that we obtain a finite, effective renormalized mass
\begin{equation}\label{R19}
    \om_{R}=Z_{2}\om=-g^{2}\left(1-\f{1}{2N}\right)\Delta_{R}
\end{equation}
with
$\De_{R}=Z_{2}\De=Z_{2}\left\langle\overline{\psi}\psi\right\rangle$
the finite, renormalized VEV of the composite operator
$\overline{\psi}\psi$.
\\\\To obtain a finite, renormalized expression for the vacuum
energy as a function of $\De_{R}$ or $\om_{R}$, we have to use the
same trick as in the unrenormalized case and consider the
renormalization of $\f{dE}{dg^{2}}$. Let us first consider the
case when the vertex $x$ hit by $\f{d}{dg^{2}}$ is a $2PPR$ vertex
and restrict ourselves to divergent subgraphs which contain $x$
(the ones not containing $x$ pose no problem and simply replace
the original $\Delta$ evaluated without counterterms by $\De$ with
counterterms included). The divergent subgraph can just end at $x$
from the left or the right (FIG.7$a$ and 7$b$) or the $2PPR$
vertex $x$ can be embedded in it (FIG.7$c$).
\begin{figure}[t]\label{fig7}
\begin{center}
    \scalebox{0.5}{\includegraphics{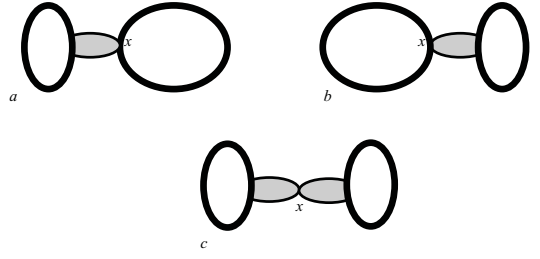}}
    \caption{Divergent subgraphs of $\f{dE}{dg^{2}}$ containing the $2PPR$ vertex $x$. }
\end{center}
\end{figure}
Graphs 7$a$ en 7$b$ can be made finite by the $2PPR$ part of the
coupling constant counterterm and making use of (\ref{R12}), their
renormalization contributes
\begin{eqnarray}\label{R20}
    (7a)+(7b)&=&2\left(-\f{1}{2}\right)\De_{kl}\de
    Z_{2;ij,kl}\left(\de_{ij}\de_{pq}-\de_{iq}\de_{pj}\right)\Delta_{pq}\nonumber\\&=&-\left(1-\f{1}{2N}\right)\de
    Z_{2}\Delta^{2}
\end{eqnarray}
where we have used (\ref{2bis}), (\ref{R12}) and (\ref{R15}).
Graph 7$c$ can be made finite with that part of coupling constant
renormalization that factorizes at the $2PPR$ vertex $x$. Its
renormalization therefore contributes
\begin{eqnarray}\label{R21}
    (7c)&=&-\f{1}{2}\De_{kl}\left(\de Z_{2;ii,kl}\de Z_{2;jj,pq}-\de Z_{2;ij,kl}\de
    Z_{2;ij,pq}\right)\De_{pq}\nonumber\\&=&-\f{1}{2}\left(1-\f{1}{2N}\right)\de
    Z_{2}^{2}\De^{2}
\end{eqnarray}
where we made use of (\ref{R15}) and (\ref{R16}). Adding the
counterterm contributions (\ref{R20}) and (\ref{R21}) to the
original unrenormalized expression (\ref{eq1}), we obtain
$-\f{1}{2}\left(1-\f{1}{2N}\right)\left(Z_{2}\De\right)^{2}=-\f{1}{2}\left(1-\f{1}{2N}\right)\De_{R}^{2}$,
which is finite. When $\f{d}{dg^{2}}$ hits a $2PPI$ vertex, we can
unambiguously subdivide the vacuum diagrams in a maximal $2PPI$
part, which contains the vertex hit by $\f{d}{dg^{2}}$, and one or
more $2PPR$ bubble insertions which, after renormalization, can be
replaced by the effective renormalized mass $\om_{R}$. We
therefore have
\begin{equation}\label{R22}
    \f{dE}{dg^{2}}=-\f{1}{2}\left(1-\f{1}{2N}\right)\De_{R}^{2}+\f{\p E_{2PPI}}{\p
    g^{2}}\left(\om_{R},g^{2}\right)
\end{equation}
We still have to show that the usual counterterms make $\f{\p
E_{2PPI}}{\p g^{2}}\left(\om_{R},g^{2}\right)$ finite. The
non-perturbative mass $\om_{R}$, running in the propagatorlines,
will now generate selfenergies which require mass renormalization,
which is not present in the original Lagrangian. Again coupling
constant renormalization will solve the problem. Let us consider a
generic selfenergy subgraph which needs mass renormalization.
Since the divergence is linear in $\om_{R}$, we can restrict
ourselves to $2PPI$ diagrams with only one $2PPR$ bubble insertion
(FIG.8).
\begin{figure}[t]\label{fig8}
\begin{center}
    \scalebox{0.5}{\includegraphics{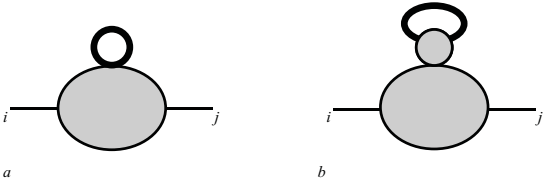}}
    \caption{Selfenergy subgraphs needing mass renormalization. }
\end{center}
\end{figure}
The divergent part of this subgraph, that one wants to
renormalize, can end at the $2PPR$ vertex (FIG.8$a$) or can
continue throughout the $2PPR$ bubble (FIG.8$b$). In the first
case, one needs the $2PPR$ part of coupling constant
renormalization which contains only one $2PPR$ vertex (because the
divergent part considered belongs to the $2PPI$ part of the
diagram). We obviously have
\begin{equation}\label{R23}
    \de
    Z_{4;ij,kl}^{2PPR,1}=\left(\de_{ij}\de_{mn}-\de_{in}\de_{mj}\right)\de
    Z_{2;mn,kl}^{2PPI}
\end{equation}
so that the counterterm contribution is
\begin{eqnarray}\label{R24}
    (8a)&=&-g^{2}\De_{kl}\de
    Z_{4;kl,ij}^{2PPR,1}\nonumber\\&=&-g^{2}\De\left(1-\f{1}{2N}\right)\de
    Z_{2}^{2PPI}\de_{ij}
\end{eqnarray}
where use was made of (\ref{R14}) and (\ref{R23}). \\In the second
case, the divergence factorizes into a $2PPR$ coupling constant
renormalization part (the bubble graph part) and a $2PPI$ mass
renormalization part, so that the counterterm contribution is
\begin{eqnarray}\label{R25}
    (8b)&=&-g^{2}\De_{kl}\de
    Z_{4;mn,kl}^{2PPR}\de Z_{2;mn,ij}^{2PPI}\nonumber\\&=&-g^{2}\De\left(1-\f{1}{2N}\right)\de
    Z_{2} \de  Z_{2}^{2PPI}\de_{ij}
\end{eqnarray}
Adding both contributions, the relevant parts of the coupling
constant counterterms give
\begin{eqnarray}\label{R26}
    (8a)+(8b)&=&-g^{2}Z_{2}\De\left(1-\f{1}{2N}\right)\de
    Z_{2}^{2PPI}\de_{ij}\nonumber\\&=&\om_{R}\de
    Z_{2}^{2PPI}\de_{ij}
\end{eqnarray}
which is exactly what we need for mass renormalization in
$E_{2PPI}$.\\\\In an analoguous way, we can consider the
logarithmic overall divergences of the vacuum diagrams which are
quadratic in $\om_{R}$. We now consider $2PPI$ vacuum diagrams
with two bubble insertions. One type of coupling constant
renormalization subgraphs end at both $2PPR$ vertices (FIG.9).
\begin{figure}[t]\label{fig9}
\begin{center}
    \scalebox{0.5}{\includegraphics{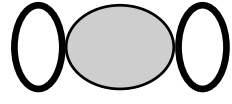}}
    \caption{Coupling constant renormalization graph with 2 $2PPR$ vertices.}
\end{center}
\end{figure}
They can be renormalized by the corresponding $2PPR$ part of the
coupling constant renormalization counterterm.
\begin{equation}\label{R27}
    \de
    Z_{4;ij,kl}^{2PPR,2}=g^{2}\left(\de_{ij}\de_{mn}-\de_{in}\de_{mj}\right)\de
    \zeta_{mn,rs}^{2PPI}\left(\de_{rs}\de_{kl}-\de_{rl}\de_{ks}\right)
\end{equation}
where $\delta \zeta_{mn,rs}$ is the overal divergent part of
$\left\langle\overline{\psi}_{m}\psi_{n}\overline{\psi}_{r}\psi_{s}\right\rangle$.
Adding the contributions from coupling constant renormalization
graphs which also go through the bubble parts, we find
\begin{equation}\label{R28}
\de E_{2PPI}=\f{1}{2}\de
    Z_{4;ij,kl}^{2PPR,2}\De_{ij}^{R}\De_{kl}^{R}=\f{1}{2}\om_{R}^{2}\delta
\zeta^{2PPI}
\end{equation}
with
\begin{equation}\label{R29}
    \delta \zeta^{2PPI}=\de \zeta_{mm,nn}^{2PPI}
\end{equation}
and use was made of (\ref{R27}).\\Again coupling constant
renormalization provides us with the necessary additive
renormalization of the $2PPI$ vacuum energy. Furthermore,
completely analogous arguments can be used to show that the
unrenormalized gap equation (\ref{eq7}) gets renormalized to
\begin{equation}\label{R30}
    \f{\p E_{2PPI}}{\p \om_{R}}\left(\om_{R},g\right)=\De_{R}
\end{equation}
\\It is clear that the $2PPI$ coupling constant and wave function
renormalization subgraphs can be renormalized with the original
counterterms. We therefore conclude that $\f{\p E_{2PPI}}{\p
g^{2}}\left(\om_{R},g\right)$ is finite and hence (\ref{R22}) is
finite and can be integrated. Making use of the gap equation
(\ref{R30}), we find
\begin{equation}\label{R31}
    E\left(g^{2}\right)=\f{1}{2}g^{2}\left(1-\f{1}{2N}\right)\De_{R}^{2}+E_{2PPI}\left(\om_{R},g^{2}\right)
\end{equation}
Of course, we also have the equivalence (\ref{tra}) in the
renormalized case.\\\\
 For the rest of the paper, it
is implicitly understood we're working with renormalized
quantities, so that we can drop the $R$-subscripts.
\section{\label{sec4}Preliminary results for the mass gap and vacuum energy}
FIG.10 shows the first terms in the loop expansion for $E_{2PPI}$.
\begin{figure}[t]\label{fig10}
\begin{center}
    \scalebox{0.6}{\includegraphics{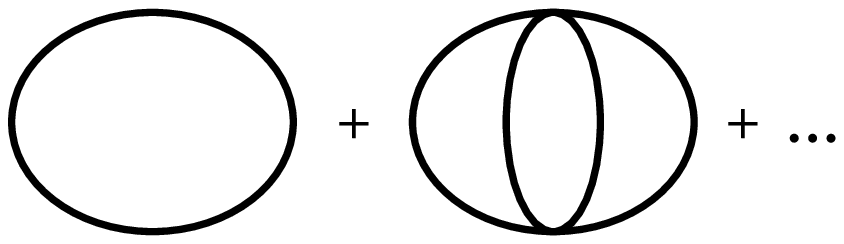}}
    \caption{$E_{2PPI}$.}
\end{center}
\end{figure}
Restricting ourselves to the 1 loop vacuum bubble, we have in
dimensional regularization with $d = 2-\varepsilon$,
\begin{equation}\label{11}
    E_{2PPI}=-2N\om^{2}\frac{1}{2-\varepsilon}\mu^{\varepsilon}\int\frac{d^{d}p}{(2\pi)^{d}}\frac{1}{p^{2}+\om^{2}}
\end{equation}
Using the $\overline{MS}$ scheme, we arrive at
\begin{equation}\label{12}
    E=\f{1}{2}g^{2}\left(1-\f{1}{2N}\right)\De^{2}+\frac{N}{4\pi}\om^{2}\left(\ln\frac{{\om}^{2}}{\omu^{2}}-1\right)
\end{equation}
The gap equation (\ref{R30}) gives
\begin{equation}\label{m1}
    \frac{N\om}{2\pi}\ln\frac{\om^{2}}{\omu^{2}}=\oD
\end{equation}
Consequently, the vacuum energy is expressed by
\begin{equation}\label{m1bis}
    E=-\frac{N}{4\pi}\om^{2}
\end{equation}
At 1 loop order, we have
\begin{equation}\label{13}
    \og^{2}(\omu)=\frac{1}{\beta_{0}\ln\frac{\omu^{2}}{\lms^{2}}}
\end{equation}
where $\beta_{0}$ is the leading order coefficient of the
$\obeta$-function
\begin{equation}\label{14}
    \omu\frac{\partial
    \og^{2}}{\partial\omu}=\obeta(\og^{2})=-2\left(\beta_{0}\og^{4}+\beta_{1}\og^{6}+\obeta_{2}\og^{8}+\cdots\right)
\end{equation}
The values of the coefficients can be found in
\cite{Gracey1,Gracey2,Gracey3,Luperini}
\begin{eqnarray}
    \label{15bis}\beta_{0}&=& \frac{N-1}{2\pi}\\
    \label{15tris}\beta_{1}&=& -\frac{N-1}{4\pi^{2}}\\
    \label{15quat}\obeta_{2}&=& -\frac{(N-1)\left(N-\frac{7}{2}\right)}{16\pi^{3}}
\end{eqnarray}
To get a numerical value for the mass gap \footnote{At 1 loop
$2PPI$ order, there is no mass renormalization, hence $\om$ is the
physical mass.}, we have to choose the subtraction scale $\omu$.
The choice immediately coming to mind is setting $\omu=\om$, which
eliminates the potentially large logarithm present in (\ref{12}).
Doing so, we find, next to the perturbative solution $\om=0$,
\begin{equation}\label{m2}
    m=\om=\lms
\end{equation}
while
\begin{equation}\label{m3}
    E=-\frac{N}{4\pi}\lms^{2}
\end{equation}
The exact mass gap is given by \cite{Forgacs}
\begin{equation}\label{17}
    m_{exact}=(4e)^{\frac{1}{2N-2}}\frac{1}{\Gamma\left(1-\frac{1}{2N-2}\right)}\lms
\end{equation}
while the exact vacuum energy is \cite{Zamo}
\begin{equation}\label{18}
    E_{exact}=-\frac{1}{8}m_{exact}^{2}\cot\left(\frac{\pi}{2(N-1)}\right)
\end{equation}
We expect that the error on $E$ consists of the error on the mass
\emph{squared} and the error on the function multiplying that mass
squared. Therefore we will consider the quantity $\sqrt{-E}$ to
test the reliability of our results. We define the deviations in
terms of percentage $P$ and $Q$, i.e.
\begin{eqnarray}
    \label{p}P &=& 100\frac{m_{eff} - m_{exact}}{m_{exact}}\\
    \label{q}Q &=& 100\frac{\sqrt{-E} - \sqrt{-E_{exact}}}{\sqrt{-E_{exact}}}
\end{eqnarray}
Looking at TABLE \ref{table1}, we notice that our results
\footnote{$Q(2)$ is not defined since $E_{exact}(2)=0$. We have
$E(2)=-0.16\lms^{2}$.} are quite acceptable.
\begin{table}
\caption{\label{table1}1 loop results for mass gap and vacuum
energy}
\begin{ruledtabular}
\begin{tabular}{ccccc}
$N$ & $P$ & $Q$ & $P_{1/N}$ & $Q_{1/N}$\\
\hline
2&-46.3\%&-&-21.9\%&-\\
3&-32.5\%&-6.7\%&-12.2\%&5.8\%\\
4&-24.2\%&-8.0\%&-7.0\%&1.3\%\\
5&-19.1\%&-7.2\%&-4.5\%&0.4\%\\
6&-15.8\%&-6.2\%&-3.1\%&0.1\%\\
7&-13.5\%&-5.5\%&-2.3\%&0.007\%\\
8&-11.7\%&-4.8\%&-1.8\%&-0.03\%\\
9&-10.4\%&-4.3\%&-1.4\%&-0.04\%\\
10&-9.3\%&-3.9\%&-1.1\%&-0.04\%\\
20&-4.6\%&-2.0\%&-0.3\%&-0.02\%\\
\end{tabular}
\end{ruledtabular}
\end{table}
We notice there is convergence ($P\rightarrow0$ and
$Q\rightarrow0$) to the exact result in case of
$N\rightarrow\infty$. In fact, we recovered the
$N\rightarrow\infty$ approximation. For comparison, we also
displayed the next to leading results
\footnote{$E_{1/N}(2)=-0.22\lms^{2}$.}, given by expanding
(\ref{17}) and (\ref{18}) in powers of $1/N$.
\begin{eqnarray}
    \label{bigN1}m_{1/N}&=&\left(1+\frac{1-\C_{E}+\ln4}{2N}+
    \mathcal{O}\left(\frac{1}{N^{2}}\right)\right)\lms\\
    \label{bigN2}E_{1/N}&=&\left(-\frac{N}{4\pi}+\frac{\C_{E}-\ln4}{4\pi}+
    \mathcal{O}\left(\frac{1}{N}\right)\right)\lms^{2}
\end{eqnarray}
where $\C_{E}\approx0.577216$ is the Euler-Mascheroni constant.
\\However, the choice $\omu=\om=\lms$ cannot satisfy us, since
we are expanding in $\og^{2}(\om)=\infty$. We may have
qualitatively good results, but for a field theory where the exact
results are unknown, $\og^{2}=\infty$ gives by no means an
indication about how trustworthy our approximations are. It is
clear we must find a better method to achieve results with the
$2PPI$ expansion.
\section{\label{sec5}Optimization and 2 loop corrections}
\subsection{\label{subsec51}Renormalization group equation for $E$}
A standard approach to get better results is the usage of the
renormalization group equation (RGE). In our approach, we first
solved the gap equation and then set $\omu=\om$. Normally, when
minimizing the effective potential $V$, one first sets $\omu=\om$,
and afterwards the RGE is used to sum leading logarithms, while
all quantities are running according to their renormalization
group equations at scale $\om$. We already mentioned $E$ cannot be
treated on equal footing with an effective potential due to the
demand that $\frac{\partial E}{\partial \om}=0$ must hold. We
first point out why this also disturbs a standard RGE improvement
of $E$.\\\\Since $E$ is the vacuum energy, it is a physical
quantity and therefore, it shouldn't depend on the subtraction
scale $\omu$. This is expressed in a formal way by means of the
RGE
\begin{equation}\label{RGE1}
    \omu\frac{dE}{d\omu}=0
\end{equation}
In a perturbative series expansion, this means the differential
equation (\ref{RGE1}) must be fulfilled order by order, when all
quantities obey their running w.r.t. $\omu$. Out of (\ref{R14}),
we extract the running of $\om$, namely
\begin{equation}\label{RGE2}
    \omu\frac{\partial \om}{\partial \omu} =\left(\frac{\obeta(\og^{2})}{\og^{2}}+\ogamma(\og^{2})\right)\om\equiv\okappa(\og^{2})\om
\end{equation}
where $\ogamma(\og^{2})$ governs the scaling behaviour of $\oD$
\begin{equation}\label{RGE3}
    \omu\frac{\partial \oD}{\partial \omu} =\ogamma(\og^{2})\oD
\end{equation}
with
\begin{equation}\label{RGE4}
    \ogamma(\og^{2})=\gamma_{0}\og^{2}+\ogamma_{1}\og^{4}+\ogamma_{2}\og^{6}+\cdots
\end{equation}
The coefficients are given by
\cite{Gracey1,Gracey2,Gracey3,Luperini}
\begin{eqnarray}
    \label{RGE4bis}\gamma_{0}&=&\frac{N-\frac{1}{2}}{\pi}\\
    \label{RGE4tris}\ogamma_{1}&=&-\frac{N-\frac{1}{2}}{4\pi^{2}}\\
    \label{RGE4quat}\ogamma_{2}&=&-\frac{\left(N-\frac{1}{2}\right)\left(N-\frac{3}{4}\right)}{4\pi^{3}}
\end{eqnarray}
After some calculation, we find
\begin{eqnarray}
    \label{RGE5}\omu\frac{dE}{d\omu}&=&\left(\omu\frac{\partial}{\partial \omu}+\obeta(\og^{2})\frac{\partial}{\partial \og^{2}}+\okappa(\og^{2})\om\frac{\partial}{\partial \om}\right)E
    \nonumber\\&=& \frac{1}{4\pi}\frac{\om^{2}}{1-\frac{1}{2N}}+\mathcal{O}(\og^{2})
\end{eqnarray}
It seems that $E$ doesn't obey its RGE. Perturbatively, it of
course fullfills the RGE up to $\mathcal{O}\left(g^{2}\right)$
since $\om=0$ to all orders in perturbation theory. We must not be
tempted to interpret this failure as the need to introduce some
non-perturbative running coupling constant, as can be found in
literature sometimes. The nature of the \emph{apparent} problem
lies in the fact that we forgot about the gap equation
$\frac{\partial E}{\partial \om}=0$, because only then our $2PPI$
expression for $E$ is meaningful. (\ref{m1}) gives that
\begin{equation}\label{RGE6}
    \ln\frac{\om^{2}}{\omu^{2}}\propto\frac{1}{\og^{2}}
\end{equation}
It is easy to check that (\ref{RGE6}) means that all leading log
terms in the expansion of $E$ are of the order unity.
Consequently, we cannot simply show order by order that $\omu\frac{dE}{d\omu}=0$.\\
The problem extends to higher orders: when we would calculate $E$
up to a certain order $n$, we would need knowledge of \emph{all}
leading, subleading,..., $n^{th}$ leading log terms.\\The above
discussion reveals a possible strategy : we could do a (leading)
log expansion for $E_{2PPI}$, with a source $J$ coupled to
$\overline{\psi}\psi$. Then we could use the RGE for $E$ to sum
all (leading) logs in $E_{2PPI}$. We leave this idea, because the
RGE for $E$ itself is non-linear when $J\neq0$ \footnote{We must
add a source term, otherwise $E$ cannot be treated as an effective
potential in the usual sense.}. This is accompanied with its own
problems. A thorough discussion of this subject can be consulted
in \cite{Verschelde3}.\\\\We conclude that we cannot use the RGE
for $E$ to optimize that what we did hitherto. The crucial point
is that the gap equation must hold for consistency. We can only
set $\omu=\om$ in $\frac{\partial E}{\partial\om}$ \emph{after}
deriving $E$ w.r.t. $\om$ and solving this gap equation, \emph{not
before}.
\subsection{\label{subsec52}Optimization}
We have seen that the $\ms$ scheme is not optimal for the $2PPI$
expansion used on GN. We could have renormalized the coupling
constant in another way and hope that this gives better results.
It is easily verified that going to a scheme with coupling
$g^{2}$, determined at lowest order by
$\og^{2}=g^{2}\left(1+b_{0}g^{2}\right)$, gives the same results
as in (\ref{m2}) and (\ref{m3}), but now with $g^{2}=b_{0}^{-1}$.
This means results are as good as before, but for a sufficiently
large $b_{0}$, $g^{2}$ is small. Again, we put $\omu=\om$ to
cancel logarithms.\\ Till now, we kept $\om$ as the mass
parameter, however we should go to another scheme for this
quantity too. The results are then no longer independent of the
renormalization prescriptions, i.e. if
$\om=m\left(1+a_{0}g^2\right)$ at lowest order, then $a_{0}$
enters the final results, and $a_{0}$ is completely free to
choose. We tackle the problem of freedom of renormalization of the
coupling constant and mass parameter in 4 consecutive steps.\\\\\\
\underline{Step 1}\\
First of all, we remove the freedom how the mass parameter is
renormalized. We can replace $\om$ by an unique \footnote{Up to an
irrelevant (integration) constant that can be dropped.} $M$ such
that $M$ is renormalization scale and  scheme independent (RSSI)
\cite{Karel}. Out of (\ref{RGE2}), we immediately deduce that
\begin{equation}\label{o1}
    M=\overline{f}(\og^{2})\om
\end{equation}
where $\overline{f}(\og^{2})$ is the solution of
\begin{equation}\label{o2}
    \omu\frac{\partial \overline{f}}{\partial
    \omu}=-\okappa(\og^{2})\overline{f}
\end{equation}
When we change our MRS, we have relations of the form
\begin{eqnarray}
    \label{o3a}\og^{2}&=&g^{2}\left(1+b_{0}g^{2}+b_{1}g^{4}+\cdots\right)\\
    \label{o3b}\om&=&m\left(1+m_{0}g^{2}+m_{1}g^{4}+\cdots\right)\\
    \label{o3c}\overline{f}(\og^{2})&=&f(g^{2})\left(1+f_{0}g^{2}+f_{1}g^{4}+\cdots\right)
\end{eqnarray}
Whenever a quantity is barred, it's understood we're considering
$\ms$, otherwise we're considering an arbitrary MRS \footnote{We
notice that $\B_{0}$, $\B_{1}$ and $\C_{0}$ are the same for each
MRS.}. Using the foregoing relations, it is easy to show
the scheme independence of $M$. \\\\
The explicit solution, up to the order we will need it, is given
by
\begin{eqnarray}\label{o4}
    \overline{f}(\og^{2})&=&(\og^{2})^{-1+\frac{\C_{0}}{2\B_{0}}}\left\{1\vphantom{\frac{\C_{0}\left(\frac{\B_{1}^{2}}{\B_{0}^{2}}-\frac{\obeta_{2}}{\B_{0}}\right)}{\B_{0}}}+\frac{\og^{2}}{2}\left(-\frac{\B_{1}\C_{0}}{\B_{0}^{2}}+\frac{\ogamma_{1}}{\B_{0}}\right)\right.\nonumber\\&+&\left.\frac{\og^{4}}{4}\left[\vphantom{\frac{\C_{0}\left(\frac{\B_{1}^{2}}{\B_{0}^{2}}-\frac{\obeta_{2}}{\B_{0}}\right)}{\B_{0}}}\frac{1}{2}\left(-\frac{\B_{1}\C_{0}}{\B_{0}^{2}}+\frac{\ogamma_{1}}{\B_{0}}\right)^{2}\right.\right.\nonumber\\&+&\left.\left.\frac{\C_{0}\left(\frac{\B_{1}^{2}}{\B_{0}^{2}}-\frac{\obeta_{2}}{\B_{0}}\right)}{\B_{0}}-\frac{\B_{1}\ogamma_{1}}{\B_{0}^{2}}+\frac{\ogamma_{2}}{\B_{0}}\right]\right\}
\end{eqnarray}
Next, we rewrite $\om$ in terms of $M$ by inverting (\ref{o1})
\begin{equation}\label{o5}
    \om=M(\og^{2})^{1-\frac{\C_{0}}{2\B_{0}}}\left(1+c_{1}\og^{2}+c_{2}\og^{4}\right)
\end{equation}
where
\begin{eqnarray}
    \label{o6a}c_{1}&=&\frac{1}{2}\left(\frac{\B_{1}\C_{0}}{\B_{0}^{2}}-\frac{\ogamma_{1}}{\B_{0}}\right)\\
    \label{o6b}c_{2}&=&\frac{1}{8}\left(-\frac{\B_{1}\C_{0}}{\B_{0}^{2}}+\frac{\ogamma_{1}}{\B_{0}}\right)^{2}-\frac{1}{4}\left(\frac{\C_{0}\left(\frac{\B_{1}^{2}}{\B_{0}^{2}}-\frac{\obeta_{2}}{\B_{0}}\right)}{\B_{0}}\right)\nonumber\\
    &+&\frac{1}{4}\left(\frac{\B_{1}\ogamma_{1}}{\B_{0}^{2}}-\frac{\ogamma_{2}}{\B_{0}}\right)
\end{eqnarray}
\\\\
\underline{Step 2}\\
Transformation (\ref{o5}) allows to rewrite $E$ in terms of $M$.
Since the next contribution to (\ref{12}) is proportional to
$\og^{4}\om^{2}$ (see FIG.10), we can rewrite $E$ up to order
$\og^{2}$ when (\ref{o5}) is applied. Explicitly,
\begin{eqnarray}\label{o7}
    E&=&M^{2}(\og^{2})^{2-\frac{\C_{0}}{\B_{0}}}\left[\vphantom{\frac{1}{2\left(1-\frac{1}{2N}\right)}\left(\frac{1}{\og^{2}}\right)}\frac{N}{4\pi}\left(1+2c_{1}\og^{2}\right)\right.\nonumber\\&\times&\left.\left(\ln\frac{M^{2}}{\omu^{2}}+\left(2-\frac{\C_{0}}{\B_{0}}\right)\ln\og^{2}\right)-\frac{N}{4\pi}\right.\nonumber\\
    &+&\left.\frac{1}{2\left(1-\frac{1}{2N}\right)}\left(\frac{1}{\og^{2}}+2c_{1}+\left(2c_{2}+c_{1}^{2}\right)\og^{2}\right)\right]
\end{eqnarray}
It is important to notice that the demand $\frac{\partial
E}{\partial \om}=0$ is translated into $\frac{\partial E}{\partial
M}=0$, because $M$ and $\om$ differ only by an overall factor
$\overline{f}$ which depends solely on $\og^{2}(\omu)$.
\\\\
\underline{Step 3}\\
(\ref{o7}) is still written in terms of $\og^{2}$. Using
(\ref{o3a}), we exchange $\og^{2}$ for $g^{2}$, where the $b_{i}$
parametrize the coupling constant renormalization. We find
\begin{eqnarray}\label{o8}
    E&=&M^{2}(g^{2})^{2-\frac{\C_{0}}{\B_{0}}}\left(\frac{e_{-1}}{g^{2}}+e_{0}+e_{1}g^{2}\right)
\end{eqnarray}
with
\begin{eqnarray}
    \label{o9a}e_{-1}&=&\frac{1}{2\left(1-\frac{1}{2N}\right)}\\
    \label{o9b}e_{0}&=&-\frac{N}{4\pi}+\frac{-b_{0}+2c_{1}}{2\left(1-\frac{1}{2N}\right)}+\frac{b_{0}U}{2\left(1-\frac{1}{2N}\right)}+\frac{N}{4\pi}V
\end{eqnarray}
\begin{eqnarray}
    \label{o9c}e_{1}&=&\frac{b_{0}^{2}-b_{1}+c_{1}^{2}+2c_{2}}{2\left(1-\frac{1}{2N}\right)}\nonumber\\
        &+&\frac{1}{2\left(1-\frac{1}{2N}\right)}\left(b_{1}U+\frac{b_{0}^{2}}{2}U(U-1)\right)\nonumber\\
        &+&b_{0}U\left(-\frac{N}{4\pi}+\frac{-b_{0}+2c_{1}}{2\left(1-\frac{1}{2N}\right)}+\frac{N}{4\pi}V\right)\nonumber\\
        &+&\frac{N}{4\pi}\left(b_{0}U+2c_{1}V\right)\\
    \label{o9d}U&=&2-\frac{\C_{0}}{\B_{0}}\\
    \label{o9e}V&=&\ln{\frac{M^{2}}{\omu^{2}}}+\left(2-\frac{\C_{0}}{\B_{0}}\right)\ln
g^{2}
\end{eqnarray}
\\\\
\underline{Step 4}\\
Consider (\ref{o8}). We notice that the degrees of freedom,
concerning the scheme, are settled in the $b_{i}$. When we rewrite
the expansion in terms of $g^{2}_{1loop}$ instead of $g^{2}$, all
scheme dependence is reduced to \emph{one} parameter, namely
$b_{0}$. This was also recognized in \cite{Karel}. It is in a way
more "natural" to rewrite a perturbative series in terms of
$g^{2}_{1loop}$, because $g^{2}$ itself is changed whenever we
include the next loop order, while $g^{2}_{1loop}$ of course
remains the same.\\ The necessary formulas are given by
\begin{eqnarray}\label{o10}
    {g}^{2}(\omu)&=&\frac{1}{x}-\frac{\beta_{1}}{\beta_{0}}\frac{\ln\frac{x}{\beta_{0}}}{x^{2}}\nonumber\\
    &+&\frac{\left(\frac{\beta_{1}}{\beta_{0}}\right)^{2}\left(\left(\ln\frac{x}{\beta_{0}}\right)^{2}-\ln\frac{x}{\beta_{0}}\right)+\left(\frac{\beta_{2}}{\beta_{0}}-\left(\frac{\beta_{1}}{\beta_{0}}\right)^{2}\right)}{x^{3}}\nonumber\\
    &+&\mathcal{O}\left(\frac{1}{x^{4}}\right)
\end{eqnarray}
where
\begin{equation}\label{o11}
    x=\frac{1}{g^{2}_{1loop}}=\beta_{0}\ln\frac{\omu^{2}}{\Lambda^{2}}
\end{equation}
$\Lambda$ is the scale parameter of the corresponding MRS. In
\cite{Celmaster}, it was shown that
\begin{equation}\label{o12}
    \Lambda=\lms e^{-\frac{b_{0}}{2\beta_{0}}}
\end{equation}
For $\beta_{2}$, we have \cite{Grunberg}
\begin{equation}\label{o13}
    \beta_{2}=(b_{0}^{2}-b_{1})\beta_{0}+\beta_{1}b_{0}+\overline{\beta}_{2}
\end{equation}
Since (\ref{o8}) is correct up to order $g^{2}M^{2}$, we can
expand up to order $x^{-1}M^{2}$.  Using (\ref{o10}), (\ref{o12})
and (\ref{o13}), the vacuum energy becomes
\begin{eqnarray}\label{o14}
    E&=&M^{2}\left(\frac{1}{x}\right)^{2-\frac{\C_{0}}{\B_{0}}}\left(E_{-1}x+E_{0}+\frac{E_{1}}{x}\right)
\end{eqnarray}
with
\begin{eqnarray}
    \label{o15a}E_{-1}&=&\frac{1}{2\left(1-\frac{1}{2N}\right)}\\
    \label{o15b}E_{0}&=&-\frac{N}{4\pi}+\frac{-b_{0}+2c_{1}}{2\left(1-\frac{1}{2N}\right)}+\frac{b_{0}U}{2\left(1-\frac{1}{2N}\right)}\nonumber\\&-&\frac{\B_{1}(U-1)L}{2\B_{0}\left(1-\frac{1}{2N}\right)}+\frac{N}{4\pi}W\\
    \label{o15c}E_{1}&=&\frac{c_{1}^{2}+2c_{2}}{2\left(1-\frac{1}{2N}\right)}+\frac{\frac{\B_{1}^{2}}{\B_{0}^{2}}(1+L)-\frac{b_{0}\B_{1}+\obeta_{2}}{\B_{0}}}{2\left(1-\frac{1}{2N}\right)}\nonumber\\
    &-&\frac{N}{4\pi}\frac{\B_{1}}{\B_{0}}UL+\frac{b_{0}^{2}U(U-1)}{4\left(1-\frac{1}{2N}\right)}+\frac{1}{2\left(1-\frac{1}{2N}\right)}\nonumber\\
    &\times&\left[\left(-\frac{\B_{1}^{2}}{\B_{0}^{2}}\left(1+L-L^{2}\right)+\frac{b_{0}^{2}\B_{0}+b_{0}\B_{1}+\obeta_{2}}{\B_{0}}\right)U\right.\nonumber\\
    &+&\left.\frac{\B_{1}^{2}L^{2}U(U-1)}{2\B_{0}^{2}}\right]+b_{0}U\left[\vphantom{\frac{2c_{1}-b_{0}}{2\left(1-\frac{1}{2N}\right)}}\frac{N}{4\pi}(W-1)\right.\nonumber\\
    &+&\left.\frac{2c_{1}-b_{0}}{2\left(1-\frac{1}{2N}\right)}\right]
    -\frac{\B_{1}}{\B_{0}}LU\left[\vphantom{\frac{2c_{1}-b_{0}}{2\left(1-\frac{1}{2N}\right)}}\frac{N}{4\pi}(W-1)\right.\nonumber\\
    &+&\left.\frac{2c_{1}-b_{0}}{2\left(1-\frac{1}{2N}\right)}+\frac{\B_{1}L}{2\B_{0}\left(1-\frac{1}{2N}\right)}
    +\frac{b_{0}U}{2\left(1-\frac{1}{2N}\right)}\right]\nonumber\\
    &+&\frac{N}{4\pi}\left(b_{0}U+2c_{1}W\right)\\
    \label{o15d}L&=&\ln\frac{x}{\B_{0}}\\
    \label{o15e}U&=&2-\frac{\C_{0}}{\B_{0}}\\
    \label{o15f}W&=&\ln{\frac{M^{2}}{\omu^{2}}}+\left(2-\frac{\C_{0}}{\B_{0}}\right)\ln\frac{1}{x}
\end{eqnarray}
\subsection{\label{subsec53}2 loop corrections}
The next order corrections are 2 loop for the mass (the setting
sun diagram of FIG.11) and 3 loop for the vacuum energy (the
basket ball diagram of FIG.2). We will restrict ourselves to 2
loop corrections. The diagram displayed in FIG.11 gives a mass
renormalization. The double line is the full propagator
$S_{full}(p)$. We first employ the $\ms$ scheme again for the
calculation.
\begin{figure}[t]\label{fig11}
\begin{center}
    \scalebox{0.6}{\includegraphics{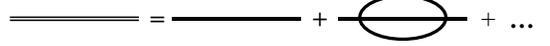}}
    \caption{Diagrams needed to calculate $m_{eff}$ in function of $\om$.}
\end{center}
\end{figure}
\\Let $\mathcal{P}$ be the value of the (amputated) setting sun
diagram. Since
\begin{equation}\label{26}
    S(p)=\frac{1}{ip\hspace{-0.125cm}/+\om}
\end{equation}
we have
\begin{equation}\label{27}
    S_{2 loop}(p)=\frac{1}{ip\hspace{-0.125cm}/+\om-\mathcal{P}}
\end{equation}
The effective mass $m_{eff}$ is the pole of $S_{2 loop}$(p). From
\cite{Verschelde3}, we obtain
\begin{eqnarray}\label{29}
    \mathcal{P}&=&\left(N-\frac{1}{2}\right)\og^{4}\left(-\om I^{2}+ip\hspace{-0.13cm}/
\frac{\varepsilon}{2-\varepsilon}I^{2}\right.\nonumber \\
&+&\left.\frac{1}{16\pi^{2}}\left(\om F_{1}+ip\hspace{-0.13cm}/
    F_{2}\right)\right)
\end{eqnarray}
where
\begin{eqnarray}
    \label{30a}I&=&\frac{1}{4\pi}\left[\frac{2}{\varepsilon}-\ln\frac{\om^{2}}{\omu^{2}}\right.\nonumber\\&+&\left.\frac{\varepsilon}{4}\left(\frac{\pi^{2}}{6}+\ln^{2}\frac{\om^{2}}{\omu^{2}}\right)+\mathcal{O}(\varepsilon^{2})\right]\\
    \label{30b}F_{1}&=&-\frac{2\pi^{2}}{9}+12q_{1}-24q_{2}\\
    \label{30c}F_{2}&=&2-\frac{2\pi^{2}}{3}\\
    \label{30d}q_{1}&=&\int_{0}^{1}dt\frac{\ln t}{t^{2}-t+1}\approx-1.17195\\
    \label{30e}q_{2}&=&\int_{0}^{1}dt\frac{\ln t}{t^{3}+1}\approx-0.951518
\end{eqnarray}
Working up to order $\og^{4}$, we find for the inverse propagator
\begin{eqnarray}\label{31}
    &&S_{2 loop}^{-1}=ip\hspace{-0.125cm}/\left[1-\frac{\left(N-\frac{1}{2}\right)\og^{4}}{16\pi^{2}}\left(1-2\ln\frac{\om^{2}}{\omu^{2}}+F_{2}\right)\right]\nonumber\\&+&\om\left[1+\frac{\left(N-\frac{1}{2}\right)\og^{4}}{16\pi^{2}}\left(2\ln^{2}\frac{\om^{2}}{\omu^{2}}+\frac{\pi^{2}}{6}-F_{1}\right)\right]
\end{eqnarray}
Solving for the pole gives
\begin{eqnarray}\label{32}
    m_{eff}&=&\om\left[1+\frac{\left(N-\frac{1}{2}\right)\og^{4}}{16\pi^{2}}\left(2\ln^{2}\frac{\om^{2}}{\omu^{2}}\right.\right.\nonumber\\&+&\left.\left.\vphantom{\frac{\left(N-\frac{1}{2}\right)\og^{4}}{16\pi^{2}}}\frac{\pi^{2}}{6}-F_{1}+1-2\ln\frac{\om^{2}}{\omu^{2}}+F_{2}\right)\right]
\end{eqnarray}
With $\omu=\om=\lms$, the above equation has no sense.\\\\Next, we
follow the same steps as executed for $E$ to reexpress $m_{eff}$
in terms of $M$ and $x$. A little algebra results in
\begin{eqnarray}\label{33}
    m_{eff}&=&M\left(\frac{1}{x}\right)^{1-\frac{\C_{0}}{2\B_{0}}}\left\{1+\frac{1}{x}\left[c_{1}+\frac{b_{0}U}{2}-\frac{\B_{1}UL}{2\B_{0}}\right]\right.\nonumber\\
    &+&\left.\frac{1}{x^{2}}\left[b_{0}c_{1}\left(1+\frac{U}{2}\right)+c_{2}+\frac{U}{2}\times\right.\right.\nonumber\\
    &&\left.\left.\left(\frac{\B_{1}^{2}}{\B_{0}^{2}}\left(L^{2}-L-1\right)+\frac{b_{0}^{2}\B_{0}+b_{0}\B_{1}+\obeta_{2}}{\B_{0}}\right)\right.\right.\nonumber\\
    &-&\left.\left.\frac{b_{0}^{2}\C_{0}U}{8\B_{0}}-\frac{L^{2}\B_{1}^{2}\C_{0}U}{8\B_{0}^{3}}-\frac{L\B_{1}\left(1+\frac{U}{2}\right)\left(c_{1}+b_{0}\frac{U}{2}\right)}{\B_{0}}\right.\right.\nonumber\\
    &+&\left.\left.\frac{N-\frac{1}{2}}{16\pi^{2}}\left(\frac{\pi^{2}}{6}+1-F_{1}+F_{2}-2W+2W^{2}\right)\right]\right\}\nonumber\\
\end{eqnarray}
The quantities $c_{1}$, $c_{2}$, $U$, $W$ and $L$ are the same as
defined before. Again, only $b_{0}$ is left over as scheme
parameter.
\section{\label{sec6}Second numerical results for the
mass gap and vacuum energy} We first discuss how we can fix the
parameter $b_{0}$ in a reasonable, \emph{self-consistent} way. A
frequently used method is the \emph{principle of minimal
sensitivity}(PMS)\cite{Stevenson}. This is based on the concept
that \emph{physical} quantities should not depend on the
renormalization prescriptions. In our case, the vacuum energy $E$
as well as the mass gap $m_{eff}$ are physical, so we could apply
PMS. However, PMS doesn't always work out. Sometimes there is no
minimum, then an alternative is picking that $b_{0}$ for which the
derivative of the considered quantity is minimal ($\rightarrow$ as
near as possible to a minimum). Also \emph{fastest apparent
convergence criteria} (FACC) can be practiced.\\\\ But maybe the
biggest barrier to a fruitful use of PMS (or FACC) arises from the
same origin why $E$ didn't seem to obey its RGE. Just as the scale
dependence of $E$ is not cancelled order by order, the scheme
dependence of $E$ won't cancel order by order, so we may find no
optimal $b_{0}$, and even if we would have such $b_{0}$, it
wouldn't be certain that the corresponding $E$ really is a good
approximation to $E_{exact}$. The same obstacle will arise for the
mass gap $m_{eff}$.\\\\Apparently, we haven't got any further. We
may have a way out through. $M$, as defined in (\ref{o1}), is
RSSI, independent of the fact that it satisfies its gap equation
or not. The $2PPI$ formalism provides us with an equation to
calculate $M$ approximately. This equation, $\frac{\partial
E}{\partial M}=0$, is correct up to a certain order and $M$ is
RSSI up to that order by construction. Hence, we can ask that the
(non-zero) solution $M$ has minimal dependence on $b_{0}$. This
also gives a value for $b_{0}$ to calculate the vacuum energy,
because the $b_{0}$ for $E$ and $M$ must be equal, again because
$E$ is only correct when the gap equation is fulfilled. Also the
mass gap $m_{eff}$ can be calculated with this $b_{0}$.
\subsection{\label{subsec61}First order results}
We start from the expression (\ref{o14}), but we first restrict
ourselves to the lowest order correction.
\begin{eqnarray}\label{o16}
    E&=&M^{2}\left(\frac{1}{x}\right)^{2-\frac{\C_{0}}{\B_{0}}}\left(E_{-1}x+E_{0}\right)
\end{eqnarray}
Until now, we haven't said anything about the freedom in scale
$\omu$. Analogously as we fixed $b_{0}$, we can ask
$\frac{\partial M}{\partial \omu}=0$ due to the scale independence
of $M$. For the sake of simplicity, we will however make a
reasonable \emph{choice} for $\omu$. In order to cancel
logarithms, we could set $\omu=M$. We refer to this as Choice I.
We observe that $\ln\frac{M^{2}}{\omu^{2}}$ always appears in the
form
$W\equiv\ln\frac{M^{2}}{\omu^{2}}+\left(2-\frac{\C_{0}}{\B_{0}}\right)\ln\frac{1}{x}$;
we could determine $\omu$ such that $W=0$, then the danger of
exploding logarithms is also averted. We refer to this as Choice
II.\\\\ TABLE \ref{table2} and TABLE \ref{table3} summarize the
corresponding results.\\
\begin{table}[t]
\caption{\label{table2}Optimized first order results for mass gap
and vacuum energy (Choice I)}
\begin{ruledtabular}
\begin{tabular}{ccccc}
$N$ & $P$ & $Q$ & $\frac{N}{4\pi x}$\\
\hline
2&?&?&?\\
3&-22.5\%&43.9\%&0.60\\
4&-19.4\%&25.9\%&0.56\\
5&-16.8\%&17.9\%&0.55\\
6&-14.6\%&13.8\%&0.54\\
7&-12.7\%&11.3\%&0.53\\
8&-11.2\%&9.5\%&0.53\\
9&-10.1\%&8.2\%&0.53\\
10&-9.1\%&7.2\%&0.52\\
20&-4.5\%&3.4\%&0.50\\
\end{tabular}
\end{ruledtabular}
\end{table}
\begin{table}[t]
\caption{\label{table3}Optimized first order results for mass gap
and vacuum energy (Choice II)}
\begin{ruledtabular}
\begin{tabular}{ccccc}
$N$ & $P$ & $Q$ & $\frac{N}{4\pi x}$\\
\hline
2&?&?&?\\
3&19.9\%&120.7\%&0.30\\
4&4.5\%&57.2\%&0.31\\
5&0.3\%&36.8\%&0.32\\
6&-1.2\%&27.0\%&0.33\\
7&-1.9\%&21.3\%&0.33\\
8&-2.1\%&17.5\%&0.34\\
9&-2.2\%&14.9\%&0.34\\
10&-2.2\%&12.9\%&0.34\\
20&-1.6\%&5.5\%&0.35\\
\end{tabular}
\end{ruledtabular}
\end{table}\\
Some remarks must be made. \\\\1) We have determined the parameter
$b_{0}$ by requiring that $\left|\frac{\partial M}{\partial
b_{0}}\right|$ is minimal \footnote{No $b_{0}$ satisfying
$\frac{\partial M}{\partial b_{0}}=0$ was found.}. In FIG.12,
$M(b_{0})$ is plotted for the case $N=5$ and Choice I. FIG.13
shows $\frac{\partial M}{\partial b_{0}}$, again for $N=5$ and
Choice I. The plots for Choice II are completely similar.\\Notice
that $\left|\frac{\partial M}{\partial b_{0}}\right|$ is
relatively small. For both choices, it tended to zero for growing
$N$, f.i. $\left|\frac{\partial M}{\partial
b_{0}}\right|\approx0.045$ for $N=10$, Choice I.
\begin{figure}[t]\label{fig12fig13}
\begin{center}
    \scalebox{1}{\includegraphics{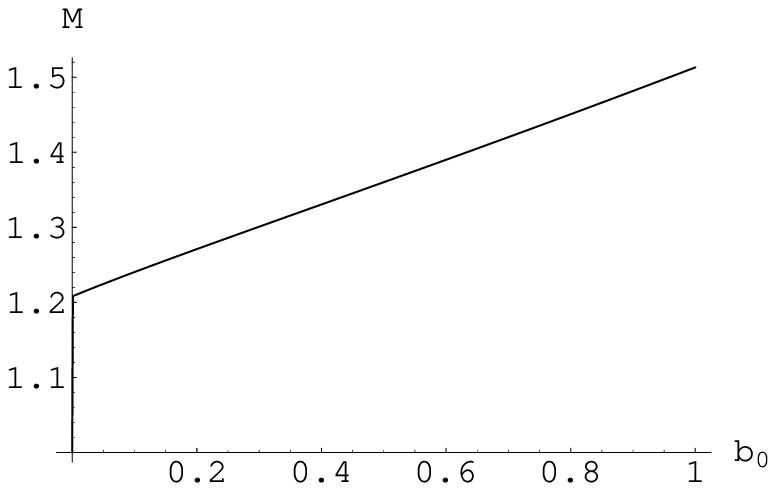}}
    \caption{$M(b_{0})$ in units of $\lms$ for $N=5$ (Choice I, 1st order).}
    \scalebox{1}{\includegraphics{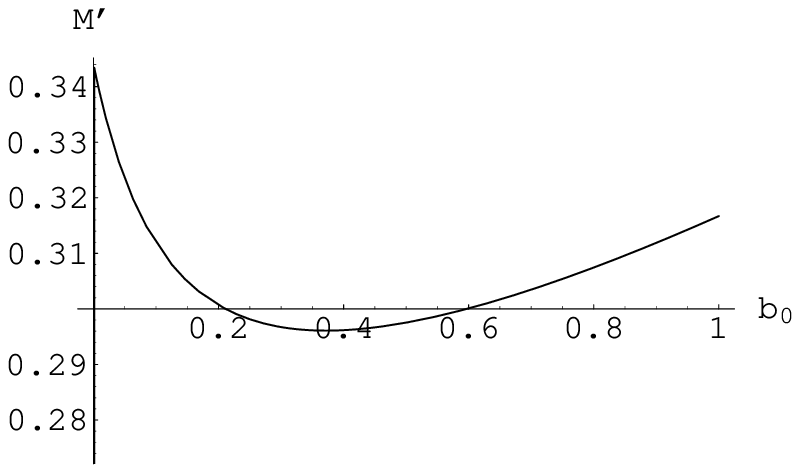}}
    \caption{$\frac{\partial M}{\partial b_{0}}$ in units of $\lms$ for $N=5$ (Choice I, 1st order).}
\end{center}
\end{figure}\\\\
Results for the mass gap agree very well with the exact values for
Choice II, this is quite remarkable since we used a lowest order
approximation.  Choice I gives almost the same results as the
$N\rightarrow\infty$ approximation. \\\\For the vacuum energy, the
results are somewhat less good than those obtained with a
straightforward $\ms$ calculation. \\Nevertheless, the mass gap as
well as the vacuum energy are converging, and we retrieve the
correct $N\rightarrow\infty$ limit. Moreover, the relevant
expansion parameter $\frac{N}{4\pi x}$ is relatively
small, and behaves more or less as a constant. \\\\
2) For $N=2$ we didn't find an optimal $b_{0}$. In the light of
the exact results (\ref{17}) and (\ref{18}), it isn't unexpected
that $N=2$ causes trouble. $N=2$ is a maximum of $m_{exact}$, and
close to $N=\frac{3}{2}$, which is a root of $m_{exact}$. There is
a sharp drop between 2 and $\frac{3}{2}$, and somewhat lower than
$\frac{3}{2}$, oscillating behaviour begins. What's more, $N=2$
and $N=\frac{3}{2}$ are both roots of $E_{exact}$, while
$E_{exact}>0$ between them. Again there is a sharp drop at
$\frac{3}{2}$ with oscillation somewhat before $\frac{3}{2}$.
Problems with $N=2$ persist at second order too, as will be seen
shortly.
\subsection{\label{subsec62}Second order results}
In TABLE \ref{table4}, we present second order results for Choice
I, while TABLE \ref{table5} displays those for Choice II. Just as
for the first order approximation, we plotted $M(b_{0})$ in
FIG.14, and $\left|\frac{\partial M}{\partial b_{0}}\right|$ in
FIG.15 for the case $N=5$, Choice I. Notice that
$\left|\frac{\partial M}{\partial b_{0}}\right|$ is smaller at
second order. For $N=10$, $\left|\frac{\partial M}{\partial
b_{0}}\right|\approx0.022$ . Again it reaches zero for infinite
$N$. Again, we weren't able to extract a value for $m_{eff}$ or
$E$ for $N=2$.
\begin{table}[t]
\caption{\label{table4}Optimized second order results for mass gap
and vacuum energy (Choice I)}
\begin{ruledtabular}
\begin{tabular}{ccccc}
$N$ & $P$ & $Q$ & $\frac{N}{4\pi x}$\\
\hline
2&?&?&?\\
3&-0.2\%&54.8\%&0.16\\
4&-2.6\%&33.5\%&0.16\\
5&-3.3\%&23.8\%&0.16\\
6&-3.7\%&18.1\%&0.16\\
7&-3.8\%&14.5\%&0.17\\
8&-3.9\%&11.9\%&0.17\\
9&-3.9\%&10.0\%&0.17\\
10&-4.0\%&8.5\%&0.17\\
20&-3.7\%&2.8\%&0.19\\
\end{tabular}
\end{ruledtabular}
\end{table}
\begin{table}[t]
\caption{\label{table5}Optimized second order results for mass gap
and vacuum energy (Choice II)}
\begin{ruledtabular}
\begin{tabular}{ccccc}
$N$ & $P$ & $Q$ & $\frac{N}{4\pi x}$\\
\hline
2&?&?&?\\
3&-4.5\%&47.7\%&0.17\\
4&-6.5\%&27.9\%&0.17\\
5&-6.1\%&19.9\%&0.17\\
6&-5.4\%&15.6\%&0.17\\
7&-4.8\%&12.8\%&0.17\\
8&-4.3\%&10.9\%&0.17\\
9&-3.9\%&9.5\%&0.17\\
10&-3.5\%&8.4\%&0.17\\
20&-1.8\%&3.9\%&0.17\\
\end{tabular}
\end{ruledtabular}
\end{table}
\subsection{\label{subsec63}Interpretation of the results}
When we compare the second with the first order results, a strange
feature immediately catches our eyes. For Choice I, the mass gap
results are better at second order, while the energy results are
worse. For Choice II, the energy results are better, while the
mass gap performs worse (except for $N=3$). To make the comparison
more transparent, we plotted the different mass gap results in
FIG.16 and energy results in FIG.17. One shouldn't be alarmed that
second order results are "worse". We see that the difference
between the Choice I and II results at first order are relatively
large, for $m_{eff}$ as well as for $E$. But at second order, the
results are almost the same for both choices, whereas
$\frac{N}{4\pi x}$ is the same. This pleases us, because these
results indicate that the choice of $\omu$ is getting less
relevant in the final results at second order. The fact that both
(reasonable) choices for the scale $\omu$ give results that are
close to each other and are converging to the same
$N\rightarrow\infty$ limit, convinces us that our method is
consistent and should give trustable results.
\begin{figure}[t]\label{fig14fig15}
\begin{center}
    \scalebox{1}{\includegraphics{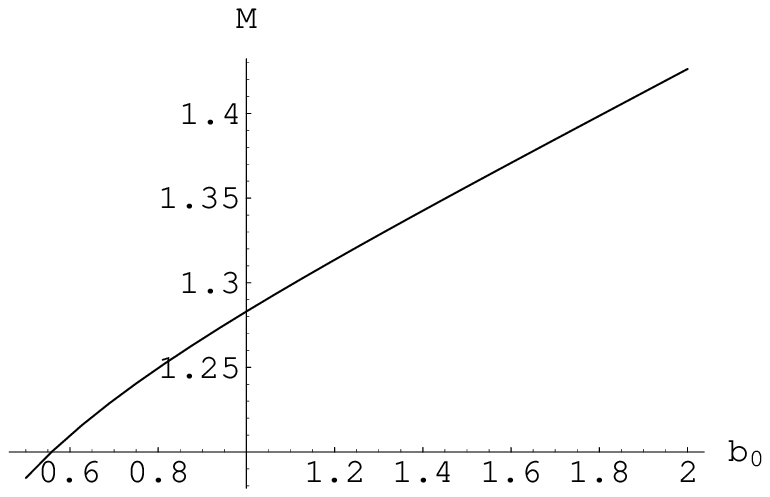}}
    \caption{$M(b_{0})$ in units of $\lms$ for $N=5$ (Choice I, 2nd order).}
    \scalebox{1}{\includegraphics{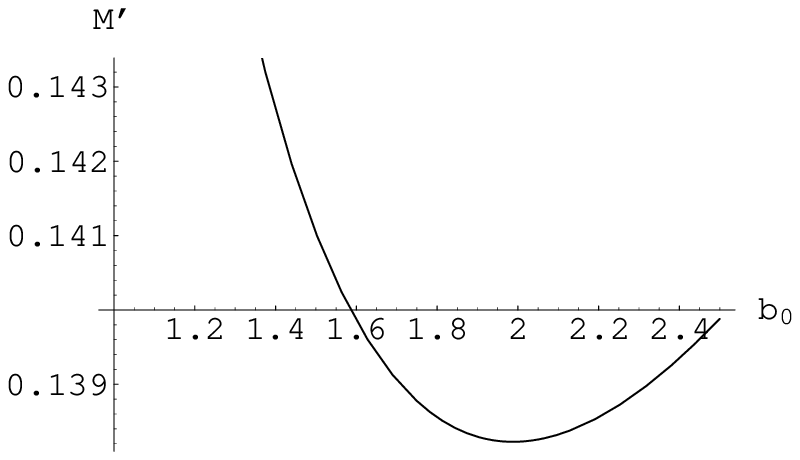}}
    \caption{$\frac{\partial M}{\partial b_{0}}$ in units of $\lms$ for $N=5$ (Choice I, 2nd order).}
\end{center}
\end{figure}
\begin{figure}[t]\label{fig16fig17}
    \begin{center}
        \scalebox{0.95}{\includegraphics{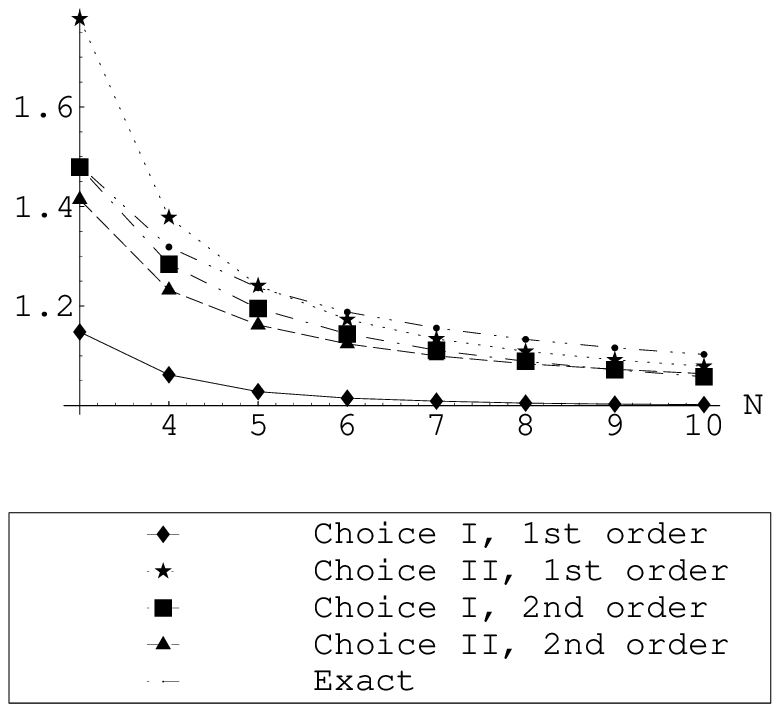}}
        \caption{Different results for $m_{eff}$.}
        \scalebox{0.95}{\includegraphics{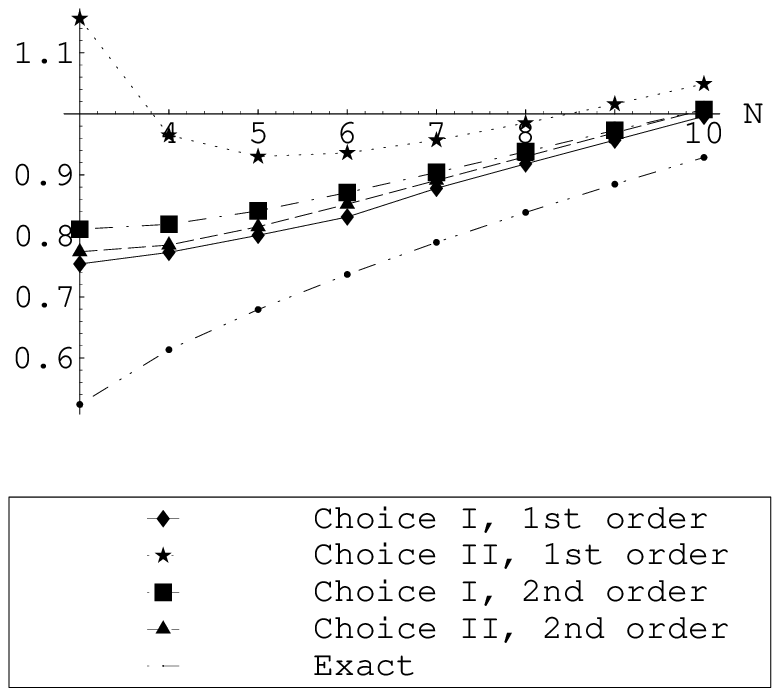}}
        \caption{Different results for $\sqrt{-E}$.}
    \end{center}
\end{figure}
\\\\Yet, there is another way to check reliability. We already said FACC
could be used as an alternative to PMS to fix $b_{0}$. More
precisely, we could use a FACC on both the energy $E$ as the mass
gap equation $\frac{\partial E}{\partial M}=0$. Explicitly, define
\begin{equation}\label{f1}
    \delta_{E}=\left|\frac{E_{1}x^{-1}-E_{0}}{E_{0}}\right|
\end{equation}
measuring the relative correction of the second order on the first
order contribution. The closer $\delta_{E}$ is to 1, the better it
is, as an indication that the series expansion is under control.
The quantity $\delta_{M}$ is defined in a similar fashion.
Unfortunately, no $b_{0}$ exists such that $\left|\frac{\partial
\delta_{E}}{\partial b_{0}}\right|$ or $\left|\frac{\partial
\delta_{M}}{\partial b_{0}}\right|$ are zero or minimal. However,
we can substitute our PMS results in $\delta_{E}$ and $\delta_{M}$
and find out what these give.\\\\Consulting FIG.18 and FIG.19, we
are able to understand why we should have ended up with
qualitatively good results, since $\delta_{E}$ as well as
$\delta_{M}$ are close to 1, even for small $N$. We also see that
both choices for $\omu$ should give comparable results, since
$\delta_{E}$ and $\delta_{M}$ fit with each other.
\\\\ We also fixed $b_{0}$ by demanding that
$\left|\frac{\partial E}{\partial b_{0}}\right|$ was minimal
\footnote{Again, no solution for $\frac{\partial E}{\partial
b_{0}}=0$.}, and we found that results were less good than those
obtained by fixing $b_{0}$ by means of $M$, except for small $N$
values \footnote{To be more precise, $Q(3)=2.1\%$ and
$Q(4)=14.4\%$. The fact that the error grows fast between $N=3$
and $N=4$, and goes slowly to $0$ for $N>5$, makes us believe it
is a rather lucky shot that the energy values are better for small
$N$.}. However, the convergence to the exact results for growing
$N$ was very slow. For example with Choice I, $Q(5)=19.4\%$,
$Q(10)=19.1\%$, $Q(20)=14.1\%$.\\An analogous story held true for
$m_{eff}$, where $b_{0}$ was determined by demanding that
$\left|\frac{\partial m_{eff}}{\partial b_{0}}\right|$ was
minimal. There, the deviation from the exact results was always
bigger \footnote{Also for $m_{eff}$, the error grows between $N=3$
and $N=5$ ($P(3)=16.8\%$, $P(4)=27.8\%$), and drops slowly to 0
for $N>5$.}, and the convergence was again rather slow. For
example, with Choice I, $P(5)=30.1\%$, $P(10)=25.7\%$,
$P(20)=18.5\%$.
\begin{figure}[t]\label{fig18fig19}
    \begin{center}
        \scalebox{0.95}{\includegraphics{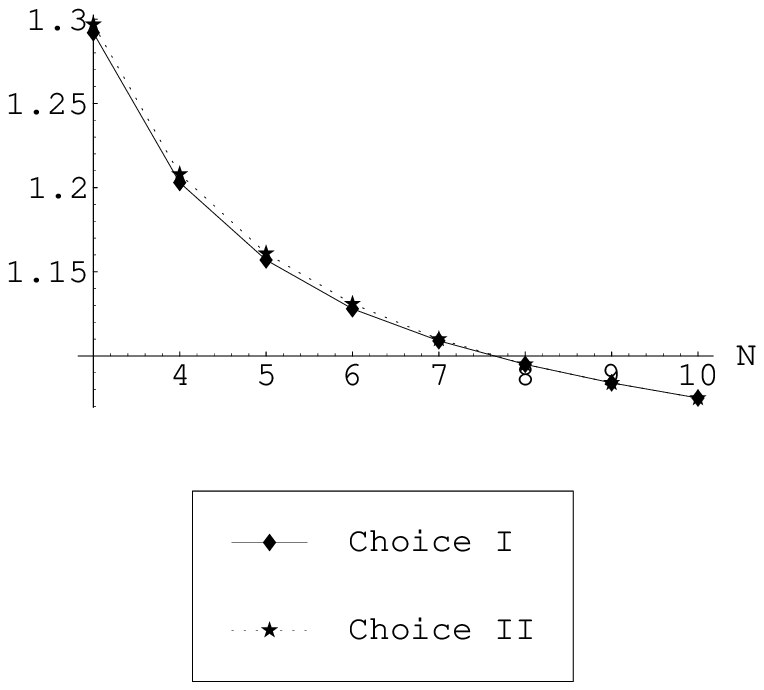}}
        \caption{$\delta_{E}$ as a function of $N$.}
        \scalebox{0.95}{\includegraphics{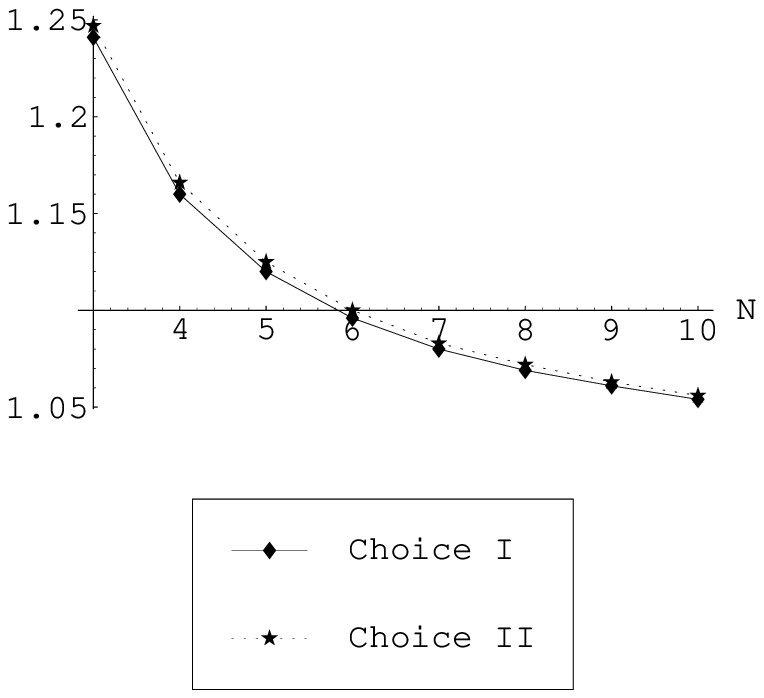}}
        \caption{$\delta_{M}$ as a function of $N$.}
    \end{center}
\end{figure}
All this corroborates our conjecture that $M$ is indeed the best
quantity to fix $b_{0}$.\\\\Before we formulate our conclusions,
we just like to mention that also in case of $N=2$ there exist a
mass gap and a non-perturbative vacuum energy. We already pointed
out why we probably didn't find an optimal $b_{0}$ with our
method. The best we can do with this special $N$ value, is just
choosing a (physical) renormalization scheme, but we must realize
we can easily obtain highly over- or underestimated values in this
case and that this is not a self-consistent way to obtain results.
\section{\label{sec7}Conclusion}
This paper, which had the purpose to investigate the dynamical
mass generation and non-perturbative vacuum energy of the
two-dimensional Gross-Neveu field theory, consisted of two main
parts. In the first part, we proved how all bubble Feynman
diagrams can be consistently resummed up to all orders in an
effective mass $\om$. We showed that this $\om$ can be calculated
from the gap equation $\frac{\partial E}{\partial \om}=0$, whereby
$E$ is the vacuum energy. $E$ is given by the sum of the $2PPI$
vacuum bubbles, calculated with the $2PPI$ massive propagator
(i.e. with mass $\om$), plus an extra term, accounting for a
double counting ambiguity.\\\\ We showed that the $2PPI$ expansion
can be renormalized with the original counterterms of the
model.\\\\A very important fact is that the $2PPI$ expansion for
$E$ is only correct if the gap equation $\frac{\partial
E}{\partial \om}=0$ is fulfilled. In this context, we discussed
the renormalization group equation for $E$, and showed why $E$
doesn't obey its RGE \emph{order by order}, because the
requirement of the gap equation turns terms of different orders
into the same order. We stress that this does not mean $E$ doesn't
obey its RGE, or ask for the introduction of a "non-perturbative"
$\beta$-function.
\\\\To get actual values for $m_{eff}$ and $E$, we employed the
$\ms$ scheme, and after the classical choice $\omu=\om$ to cancel
logarithms, we recovered the $N\rightarrow\infty$ results.
However, the corresponding coupling constant was infinite, so we
couldn't say anything about validity of the results, without the
foreknowledge of exact values. This, combined with the uselessness
of the RGE for $E$ to improve calculations, compelled us to search
for a more sophisticated way to improve the $2PPI$ technique.\\\\
In the second part, we  first eliminated the freedom in the
renormalization of the $2PPI$ mass parameter, by transforming
$\om$ to a renormalization scheme and scale independent $M$. The
consistency relation $\frac{\partial E}{\partial \om}=0$ was
completely equivalent to $\frac{\partial E}{\partial M}=0$.
Secondly, we parametrized the coupling constant renormalization.
After a reorganization of the series, all scheme dependence was
reduced to a single parameter $b_{0}$, equivalent to the choice of
a certain scale parameter $\Lambda$.\\ We fixed this $b_{0}$ by
means of the principe of minimal sensitivity (PMS). Originally,
PMS was founded on the logical requirement that observable physics
cannot depend on how one chooses to renormalize. Translated to our
case, $E$ and $m_{eff}$ shouldn't depend on the arbitrary
parameter $b_{0}$. But we showed on theoretical grounds why
applying PMS on neither $m_{eff}$ nor $E$ would be valid, because
analogously as $E$ ($m_{eff}$) doesn't lose its scale dependence
order by order, it doesn't lose its scheme dependence order by
order. \\\\Nevertheless, we gave an outcome to the problem of PMS.
By construction, $M$ is scheme and scale independent, so we can
apply PMS on this mass parameter. This provides us with an optimal
$b_{0}$ to calculate $M$, and consequently $E$ and $m_{eff}$. For
the scale $\omu$, we made 2 reasonable choices. These 2 choices
gave acceptable results at first order, yet there was quite a big
difference between them. The second order results were comparable
and qualitatively good, converging to the exact values for growing
$N$. \\\\The relevant expansion parameter was relatively small. We
gave extra evidence why results were good, by using a fastest
apparent convergence argument.\\\\We explicitly checked that using
PMS on $E$ and $m_{eff}$ to fix $b_{0}$ gave worse results, and
the convergence was very slow. \\\\Summarizing, we have
constructed a self consistent method to calculate the mass gap and
non-perturbative vacuum energy. The $2PPI$ expansion, as well as
the optimization procedure, are immediately generalizable to other
field theories.

\end{document}